\documentclass[prb,twocolumn,showpacs,preprintnumbers,amsmath,amssymb]{revtex4}
\usepackage{dcolumn}
\usepackage{bm,amsmath,amssymb}
\usepackage{graphicx}
\usepackage{bm}
\begin{document}
\title{Ferromagnetic ground state of the robust charge-ordered manganite Pr$_{0.5}$Ca$_{0.5}$MnO$_3$ obtained by minimal Al-substitution}
\author{A. Banerjee, K. Mukherjee, Kranti Kumar and P. Chaddah}
\affiliation{UGC-DAE Consortium for Scientific Research (CSR),\\ University Campus, Khandwa Road, Indore, 452 017, INDIA}
\date{\today}
\begin{abstract}
We show that minimal disturbance to the robust charge ordered Pr$_{0.5}$Ca$_{0.5}$MnO$_3$ by 2.5\% Al substitution on Mn-site drives the system towards ferromagnetic ground state. The history-dependent coexisting phases observed are explained as an outcome of a hindered first order transition with glass like arrest of kinetics resulting in irreversibility. Consistent with a simple phase diagram having ferromagnetic ground state, it is experimentally shown that these coexisting phases are far from the equilibrium. 
\end{abstract}
\maketitle
\section{INTRODUCTION} 
Perovskite manganites have become exemplary strongly correlated electron systems not only for `colossal magnetoresistence' (CMR) but also because the charge, orbital and spin degrees of freedom can be manipulated over a spectacular range by changing the band filling through doping and by changing the one-electron bandwidth through lattice distortion \cite{Tokura1}. The half doped systems play a pivotal role in the scheme of manganites assimilating several hitherto unexplained distinctive features and are under intense scrutiny \cite{Little, TVR}. The narrow bandwidth systems like Pr$_{0.5}$Ca$_{0.5}$MnO$_3$ (PCMO) or Nd$_{0.5}$Ca$_{0.5}$MnO$_3$ (NCMO) are rather special with `robust' charge and orbital ordering (CO) and CE-type spin ordering resulting in insulating (I) and antiferromagnetic (AF) ground state. This CO state can be destabilized by external stimuli or internal disorder giving rise to ferromagnetic (FM) - metallic (M) state causing colossal change of resistivity by several orders in absolute value \cite{Tokura1, Tokura2}. It is rather remarkable that the CO AF-I phase and the contrasting FM-M phase is found to be very close in energy \cite{Dag2}. That is the reason the CO state of pristine PCMO or NCMO can be `melted' by magnetic field, albeit at fields in excess to the normally available static fields ($>$200 kOe) \cite{Tokura1, Tokura2}. Significantly, CO state can be weakened and coexisting FM-M phase fraction can be easily created by introducing appropriate substitutional disorder in Mn-site \cite{Raveau1}; raising fundamental issues on the simultaneous presence of contrasting phases with very different order parameters \cite{Little}.  

It has been shown from X-ray, neutron and electron scattering that destabilization of the CO by substituting 3\% Ga in PCMO gives rise to complex field-dependent structure and magnetic order at low temperature \cite{Yaicle}. The Pr$_{0.5}$Ca$_{0.5}$Mn$_{0.97}$Ga$_{0.03}$O$_3$ is found to be in phase-separated state at low temperature. Evidence of a strained phase presumably arising from imperfect CO is found in this system. Whereas, small-angle neutron scattering has indicated the presence of micron-size ferromagnetic clusters. It is interesting to note that, recently, a model has been proposed to explain the presence of such micron size inhomogeneities in manganites from the internal strain in the system \cite{Ahn}. To understand the exact nature of the destabilized CO state we create minor perturbation in a prototype CO system. Following our earlier study \cite{Sunil1}, we substitute 2.5\% Al in Mn-site of PCMO since it has insignificant effect on structure of the pristine compound \cite{Sunil2}. It has been shown that the CO gets weakened in Pr$_{0.5}$Ca$_{0.5}$Mn$_{0.975}$Al$_{0.025}$O$_3$ but Al cannot introduce additional FM interaction unlike Co or Cr \cite {Raveau1, Tokura3, Mahe, Shoji}, since it does not have d-orbital to interact with the Mn d-band. In this letter, we show that the `robust' CO AF-I ground state is only marginally stable against small disturbance to the magnetic lattice. Through detailed measurements of history as well as time-dependent magnetization and resistivity we show that minimal Al-substitution in PCMO does not just weaken or destabilize the CO state; the ground state actually becomes FM-M.

\section{EXPERIMENTAL DETAILS} 
In this study, we have used the same Pr$_{0.5}$Ca$_{0.5}$Mn$_{0.975}$Al$_{0.025}$O$_3$ sample of Refs. \cite{Sunil1, Sunil2}. Resitivity and magnetization measurements are carried out in Quantum Design 14 Tesla PPMS-Vibrating Sample Magnetometer system.	

\section{RESULTS AND DISCUSSION} 
Figs. 1(a) and 1(b) show the magnetization(M) and resistivity (R) as a function of temperature in 40 kOe for zero-field cooled (ZFC), field-cooled cooling (FCC) and field-cooled warming (FCW) branches. Fig. 1(b) also shows the resistivity at zero field showing insulating behavior for the whole temperature range.  In Fig. 1, while hysteresis between the FCC and FCW curves in both M-T and R-T in the temperature range of about 150-30K indicates a broad first-order phase transition (FOPT) from AF-I to FM-M state, the divergence of the ZFC curve below this temperature range at the same field shows the irreversibility in the transition and points toward the non-ergodicity at low temperature. At the outset, we rule out the possibility of this state arising from spin-glass like freezing from our observation (not shown) that unlike spin-glasses the irreversibility increases with increase in measuring field.  For the same reason it cannot be ascribed to the superparamagnet like thermal blocking of clusters found in the low-field studies \cite{Sunil1}, which may only serve as the nucleating centers for the AF-I to FM-M FOPT. The field induced AF-I to FM-M transition in both M-H and R-H at 5K after cooling in zero-field is shown in the insets of Fig. 1(a) and 1(b). The irreversibility in the transition is clear from the fact that the virgin (ZFC) M-H or R-H is not traced in subsequent field cycling. All these field increasing or decreasing M-H or R-H overlap with the return cycle showing a stable FM-M phase with technical saturation around 15 kOe and negligible coercivity. The magnetization observed at 140 kOe is close to 3.5 $\mu$$_B$/formula unit. Similar magnetization value is observed for Co or Cr substitution, which is considered to be saturation moment (classical value) arising from fully aligned Mn spins \cite{Raveau1,Mahe}. Comparable magnetization values have been obtained for substitution of cations with partially or completely filled d-obitals \cite{Raveau1,Mahe}. Significantly, this FM-M state is a stable one, remains unchanged even after repeated field cycling.   

It is noteworthy that the M-T and R-T behaviors of Figs. 1(a) and 1(b) mimic the observations for La$_{5/8-x}$Pr$_x$Ca$_{3/8}$MnO$_3$ with x$\approx$0.41 (LPCMO) \cite{Sharma}, which is unambiguously shown to have FM ground state \cite{Kumar}. The non-ergodic behavior, at low temperature, which underlie the FOPT becomes clear in Figs. 1(c) and 1(d).  They show the M-T and R-T behaviors at 40 kOe after cooling the system from 320 K to 5 K in different fields (along with the ZFC, FCC and FCW of 40 kOe). It is obvious that cooling in lower fields result in lower magnetization or higher resistivity at 5 K in the same measuring field (40 kOe) because of the increasing amount of trapped AF-I phases. It is interesting to note that, when the temperature is increased from 5 K the M-T as also the R-T behaviors for curves with cooling fields higher or lower than the measuring field (40 kOe) are different. While the curves with cooling fields $<$40 kOe remain almost the same till they approach the ZFC curve, the curves with cooling fields $>$40 kOe remain distinct and finally merge with the FCW curve. The magnetization behaviors of the lower field curves are similar to the observations of Ref. \cite{Kumar}, where it is clearly shown that the arrested metastable phase gets de-arrested and transforms to equilibrium FM state with the increase in temperature.  Here we have gone further and shown through both M-T and R-T that cooling in fields higher than the measurement field traps lesser fraction of high temperature AF-I phase giving rise to higher M or lower R. The higher fractions of FM-M phase, even more than the FCC value at 40 kOe transform to AF-I phase only on approaching the superheating spinodal and merge with the FCW curve on increasing temperature. Thus Fig. 1 represents a field and temperature induced broad FOPT having hysteresis but is irreversible below $\approx$30 K. The irreversibility arises from the hindrance or critically slow dynamics of the first order transformation akin to the glass-like freezing of a fraction of the high temperature long-range order AF-I phase.   

To comprehend these apparently anomalous field-temperature induced transition and visualize various consequences, we present a schematic phase diagram in the H-T plane (Fig. 2) for this system which is in line with the projection of Ref. \cite{Kumar}. Since such transitions occur over a broad range of H and T the supercooling and superheating limits are represented by (H*, T*) and (H**, T**) bands respectively. The (H*, T*) band stretches to higher H at higher T because higher fraction of FM-M phase exists for higher H and larger T range. Similarly, the (H**, T**) band stretches to higher H at higher T. This is consistent with the expectation that at higher H, the FM phase should exist over larger temperature range. The phenomenon of kinetic arrest is represented by the (H$_K$, T$_K$) band considering the fact that the de-arrest of AF-I phase fraction to FM-M takes place over a broad H-T range. This band starts above the (H*, T*) band at zero field because ZFC gives fully arrested AF-I phase and cross the (H*, T*) band at higher H to be consistent with the observation that cooling in higher field result in transformation of larger fraction of AF-I phase to FM-M phase. The schematic in Fig. 2 shows that when the system is cooled in a field higher than that represented by the horizontal dashed line (at H$_1$), the AF-I state completely transforms to FM-M on crossing (H*, T*) band. This phase neither changes on further reduction in T nor with the decrease or increase in H indicating such paths lead to equilibrium FM-M state. It is rather interesting that whichever way one approaches this inverted triangular region between the (H*, T*) and (H$_K$, T$_K$) band, the resulting state remain stable FM-M. It is to be noted that, cooling in a field of 120 kOe gives magnetic moment close to 3.5 $\mu$$_B$/f.u. at low temperature which is expected for full saturation of the Mn moment for this sample. On the contrary, if the sample is cooled in a field below the dashed line (at H$_1$) but above the dotted line (at H$_2$), before it could cross the (H*, T*) band and completely transforms to the FM-M phase it encounters the (H$_K$, T$_K$) band and the remaining AF-I phase get arrested. This results in a metastable system with coexistence of partly transformed FM-M phase and remaining arrested AF-I phase. Cooling in a field lower than H$_2$, gives rise to completely arrested metastable AF-I phase at the lowest temperature. This kinetically arrested (or glass like) metastable state is different from the metastable supercooled state because the latter will undergo a metastable to stable transformation on lowering the temperature but not the former. In other words, relaxation time decreases with decrease in temperature for the supercooled state whereas it increases with decrease in temperature for the kinetically arrested phase \cite{Chat}.

Considering the fact that the coexisting phases arise (and persist till the lowest temperature) from the competing effects of (H*, T*) and (H$_K$, T$_K$) bands, it is rather trivial exercise to explain the observations of Fig. 1. We do further test of the phase diagram for the Pr$_{0.5}$Ca$_{0.5}$Mn$_{0.975}$Al$_{0.025}$O$_3$ sample from field annealing experiments traversing some designed paths in the H-T plane. Fig. 3(a) and 3(b) shows the M-H after cooling the sample to 5 or 25 K in different fields, then measure magnetization while isothermally reducing the field from +H to -H. Cooling in different fields renders coexisting phases with different phase fractions at the same temperatures and reduction in field has no effect on it because the AF-I phase fraction which gets arrested by the (H$_K$, T$_K$) band is not released to FM-M phase since we remain below (H$_K$, T$_K$) band while lowering field. This is obvious from the negative field cycle which show that the magnitude of magnetization remains almost same as the starting value. It is remarkable that these features are same as the observations made at 5 K for the case when Cr is substituted at Mn-site in NCMO \cite{Tokura3}. This was attributed to the formation of the FM-M domains arising from the quenched random field of Cr which contributes to the one-electron bandwidth unlike Al.  Additionally, we show in Fig. 3(c) and 3(d) the isothermal M-H after cooling the sample to 5 or 25 K in different field including zero field, then measure magnetization while H is increased. Different amount of arrested phase has resulted while traversing different amount of (H$_K$, T$_K$) and (H*, T*) bands while cooling in different fields. While increasing field, sharp change in M takes place only when the corresponding points in the (H$_K$, T$_K$) band are approached in opposite sense and de-arrest starts \cite{Kumar} . Such points in (H$_K$, T$_K$) band can be approached at lower H for higher T resulting in the striking difference in the data of 5 K and 25 K for the given slope of this band. It is interesting that similar de-arrest in field increasing cycle is observed in LPCMO \cite{Kumar}. Significantly, Figs. 3(a) and 3(b) unambiguously show that different fractions of FM phase acquired during cooling in different fields remain stable during the field reducing cycle. That is the reason, they show up with about same distinct magnitude in the -ve field increasing cycle. Moreover, if the whole sample becomes ferromagnetic at low temperature then it remains so even if the field is reduced to zero as shown in the inset of Fig. 1(a). This is emphatically shown in the resistivity measurement [inset of Fig. 1(b)] where the low resistivity value of the FM-M phase remain intact in zero field, even after repeated isothermal field cycling. This clearly suggest that, for this sample, the homogeneous FM-M phase is the stable phase at low temperature even in zero field.       

Thus we have shown that a small disturbance to robust CO state by minimal Al substitution results in field history dependent coexisting fractions of equilibrium FM-M and arrested AF-I phases. This is the manifestation of a hindered (irreversible) FOPT giving rise to glass-like kinetically arrested AF-I phase fraction at low temperatures which depends on field and can convert to the equilibrium FM-M phase. Now we give conclusive evidence to show that the non-equilibrium AF-I phase has a tendency to decay toward the FM-M state and the bulk system tends to achieve the stable FM-M phase at low temperature. It is generally considered that, for a metastable magnetic system, the field-cooled state  is the equilibrium state for that field and temperature. However, Fig. 4(a) shows the magnetization as a function of time after cooling the sample from 320 K in 80 kOe and holding the same field at 12 K. It clearly shows that magnetization which is close but less than the saturation value increases with time indicating that the sample strive to achieve fully ferromagnetic state. Moreover, it is rather drastic that when the field is reduced from 80 kOe to 40 kOe keeping the temperature constant, the magnetization still shows monotonic increase with time for many hours. Similar increase in magnetization is observed at 12 K after cooling in 50 kOe and also after reducing the field from 50 kOe to 40 kOe. The same effect becomes much more pronounced in the resistivity measurement because unlike magnetization this change is more than a factor of 10$^6$ and is non-linear with respect to the FM-M and AF-I phase fractions arising from the percolative nature. Fig. 4(b) shows the decrease in resistivity with time at different fixed temperatures after cooling the sample in 40 kOe. Similar decrease in resistivity is also observed at different stable temperatures after cooling in 80 kOe (not shown). The relaxation of resistivity is slow at lower temperatures because of the critical slowing down of the transformation process arising from the kinetic arrest. With increase in temperature the relaxation becomes faster but as the temperature approaches the (H**, T**) band the reverse transformation to AF-I state slows down the relaxation and finally its sign changes. We have shown here representative data only in a few fields and temperatures. Extensive data with detailed analysis will be communicated separately.  

\section{CONCLUSIONS} 
We conclusively show that the AF-I state of prototype robust CO system, Pr$_{0.5}$Ca$_{0.5}$MnO$_3$, is actually marginally stable and a small perturbation to its magnetic lattice can drastically change the low temperature (ground) state of the system. The observation of the history and time dependent coexisting phase at low temperature is the outcome of a first order transition which is hindered by critically slow dynamics of the transformation process similar to glass like freezing.  Significantly, our results clearly brings out the fact that the coexisting phase is not in thermodynamical equilibrium. The low temperature thermodynamic state is a homogeneous FM-M state that is masked by the kinetic arrest. The system can achieve this state by traversing along a path in high enough field, otherwise it strives to asymptotically approach this state even in low fields. Further, contrary to the earlier belief, small amount of substitution with cations without d-orbital to the magnetic lattice of PCMO can cause ferromagnetism with fully aligned spin moment of Mn. This is rather intriguing and should prompt further scrutiny of the systems where earlier studies on magnetic substitutions (like Cr, Co etc.) have been shown to introduce FM-M phase fractions in the PCMO. Moreover, this study should initiate a debate about the existence of a critical concentration and possibility of bi-critical phase competition by Mn-site substitution \cite {Tokura1}. Though at present, it is difficult to ascertain the microscopic origin of kinetic arrest resulting in glass like  critically slow dynamics of first-order transformation process, quenched disorder including strain might be playing a decisive role. Recently, a relation between the kinetic arrest and supercooling is established for disorder broadened first-order transition from phenomenological studies \cite {Kumar}. The present study should initiate a search for the microscopic origin of the kinetic arrest using various in-field microscopic as well as mesoscopic probes to establish its relation with various intrinsic parameters of the system.        

\section*{ACKNOWLEDGEMENTS} 
We are grateful to S.B. Roy for critical reading of the manuscript and thank R. Rawat for discussions. DST, Government of India is acknowledged for funding the VSM. KM acknowledges CSIR, India.

\newpage
\begin{figure*}
	\centering
		\includegraphics{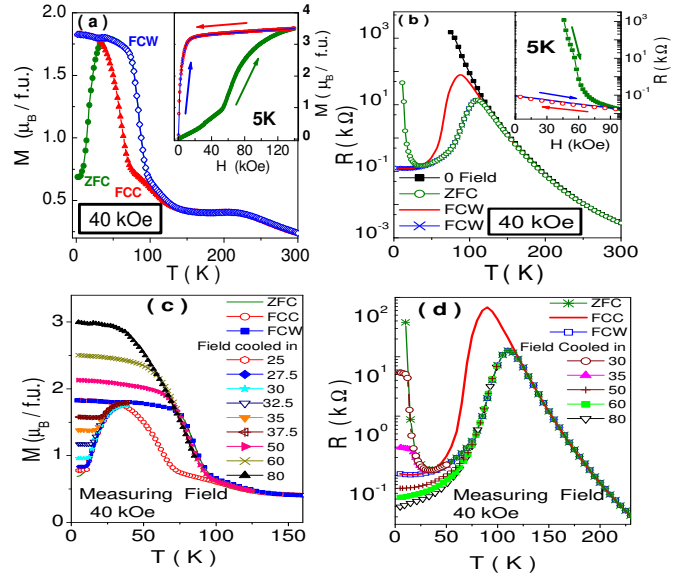}
	\caption{(Color online) History dependent M-T and R-T behaviors for Pr$_{0.5}$Ca$_{0.5}$Mn$_{0.975}$Al$_{0.025}$O$_3$ are shown. The ZFC, FCC and FCW branches of M-T and R-T at 40 kOe are shown in (a) and (b) respectively. (b) also shows the R-T in zero field. Insets of (a) and (b) show the M-H and R-H respectively, at 5 K. (c) and (d) show the M-T and R-T respectively at 40 kOe while heating, after cooling the sample from 320 K to 5 K in various indicated fields, along with the ZFC, FCC and FCW branches of 40 kOe. Different values of M and R at 5K in the same measuring field indicate different amount of arrested AF-I phase.}
	\label{fig:Fig1}
\end{figure*} 

\begin{figure*}
	\centering
		\includegraphics{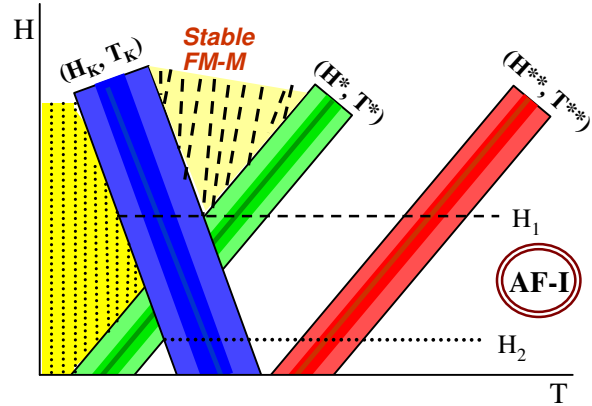}
	\caption{(Color online) Schematic H-T phase diagram for Pr$_{0.5}$Ca$_{0.5}$Mn$_{0.975}$Al$_{0.025}$O$_3$ with the  supercooling (H*, T*), superheating (H**, T**) and kinetic arrest (H$_K$, T$_K$) bands. The high temperature phase is AF-I.  The hatched inverted triangular region is a unique region in this phase diagram. The system achieves stable FM-M phase on reaching this place and it remains so on further excursions into the dotted region. Even though the equilibrium phase in the dotted region at low-T is FM-M, however, the observed phase in this region depends on the cooling field. Cooling in fields above H$_1$ give rise to stable FM-M phase once the (H*, T*) band is crossed. Cooling in fields below H$_2$ give rise to arrested AF-I phase once the (H$_K$, T$_K$) band is crossed. Cooling in fields between H$_1$ and H$_2$ result in coexisting metastable AF-I and equilibrium FM-M phase at low temperature.  }
	\label{fig:Fig2}
\end{figure*} 

\begin{figure*}
	\centering
		\includegraphics{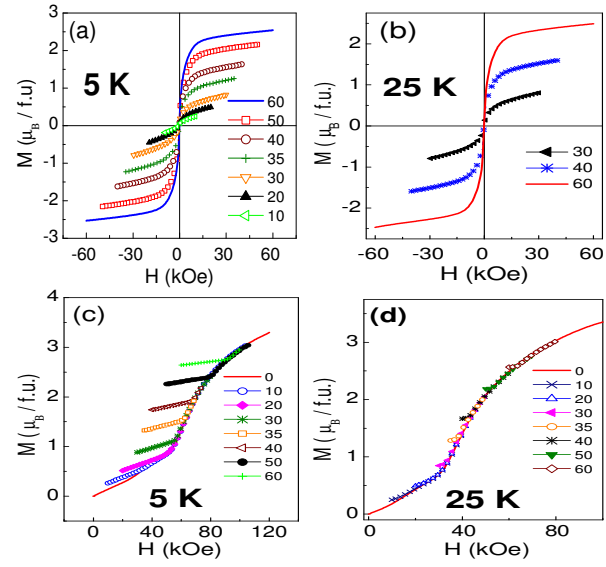}
	\caption{(Color online)  M-H of Pr$_{0.5}$Ca$_{0.5}$Mn$_{0.975}$Al$_{0.025}$O$_3$ sample after field annealing in different fields. (a) and (b) show the M-H at 5 and 25 K respectively after cooling the samples in different fields to measurement temperatures then reducing the field from +H to –H. (c) and (d) show the M-H at 5 and 25 K respectively after cooling the samples in different fields to measurement temperatures then increasing the field from the cooling field values.}
	\label{fig:Fig3}
\end{figure*} 

\begin{figure*}
	\centering
		\includegraphics{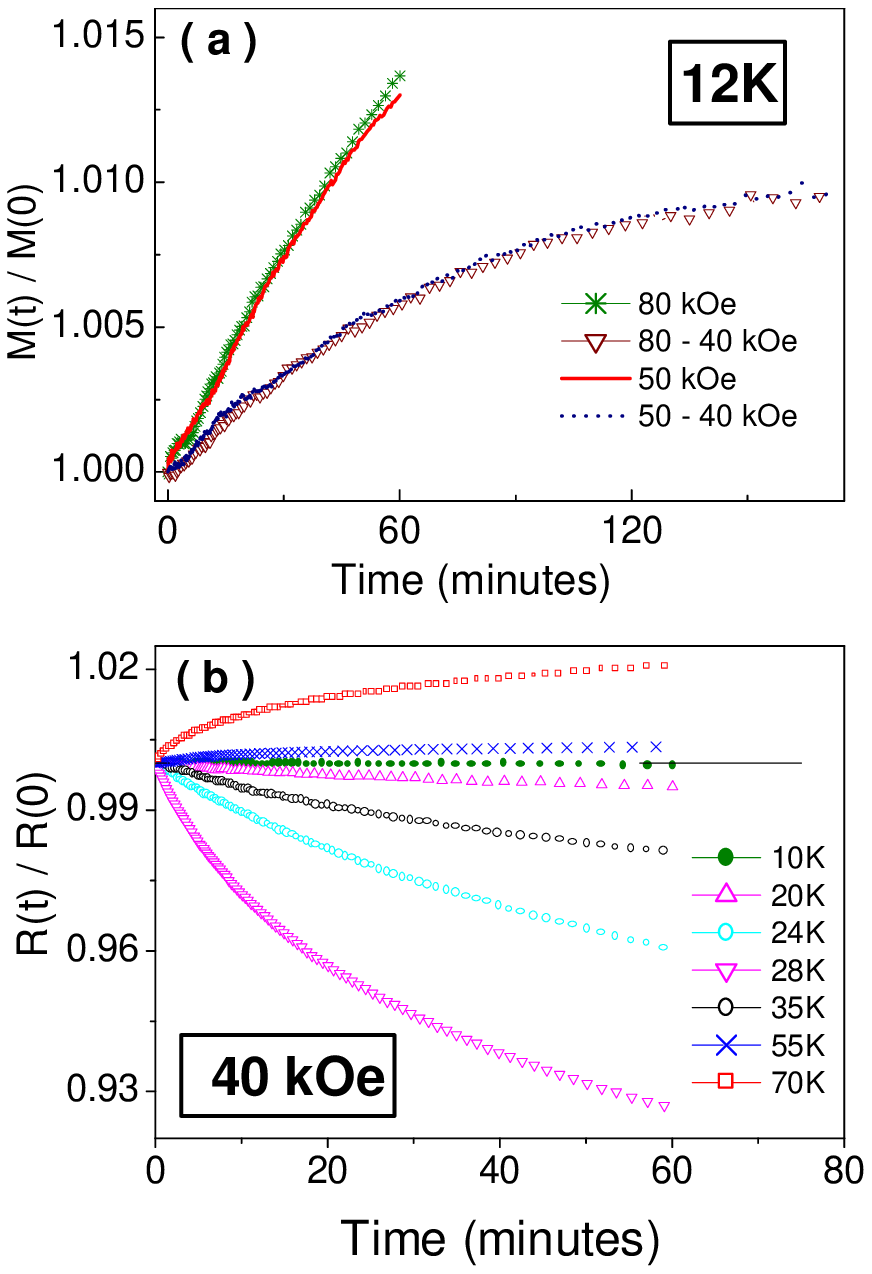}
	\caption{(Color online) Time dependence of M(t) and R(t) at constant fields and temperatures. (a) shows the M(t) normalized with respect to M (t=0) as a function of time (t) at 12 K after cooling from 320 K in 80 kOe (*) and 50 (-) kOe for an hour. Then the field is reduced to 40 kOe in both the cases [80 to 40 kOe ($\nabla$) and 50 to 40 kOe (--)] and M(t) is measured for  about six hours. After cooling from 320 K to 10 K in 40 kOe the R(t) is measured for an hour at different temperature. (b) shows the R(t) normalized with respect to R (t=0) for the respective temperatures as a function of time (t).}
	\label{fig:Fig4}
\end{figure*}


\begin{thebibliography}{ }
\bibitem[1]{Tokura1}Y. Tokura, Rep. Prog. Phys.{\bf69}, 797 (2006).
\bibitem[2]{Little}G. C. Milward, M. J. Calder\'on, and P. B. Littlewood, Nature {\bf433}, 607(2005).
\bibitem[3]{TVR}O. C\'epas, H.R. Krishnamurthy and T.V. Ramakrishnan,  Phys. Rev. Lett. {\bf94}, 247207 (2005). 
\bibitem[4]{Tokura2}Y. Tomioka, A. Asamitsu, H. Kuwahara and Y. Tokura, Phys. Rev. B {\bf53}, R1689 (1996); M. Tokunaga, N. Miura, Y. Tomioka and Y. Tokura, Phys. Rev. B {\bf57}, 5259 (1998).
\bibitem[5]{Dag2}T. Hotta and E. Dagotto, Phys. Rev. B {\bf53}, R11879 (2000).
\bibitem[6]{Raveau1}B. Raveau, A. Maignan, C. Martin and M. Hervieu, J. Solid State Chem. {\bf130}, 162 (1997); F. Damay, C. Martin, A. Maignan, M. Hervieu, B. Raveau, F. Bour\'ee and G. Andr\'e, Appl. Phys. Lett. {\bf73}, 3772 (1998); C. Martin, A. Maignan, F. Damay, M. Hervieu, B. Raveau, Z. Jirak, G. Andr\'e and F. Bour\'ee, J. Magn. Magn. Mater. {\bf202}, 11 (1999); S. H\'ebert, A. Maignan, C. Martin and B. Raveau, Solid State Commun. {\bf121}, 229 (2002); S. H\'ebert, A. Maignan, V. Hardy, C. Martin, M. Hervieu and B. Raveau,\textit{ibid.}, {\bf122}, 335 (2002); V. Hardy, A. Maignan, S. H\'ebert, C. Yaicle, C. Martin, M. Hervieu, M. R. Lees, G. Rowlands, D. Mc K. Paul and B. Raveau, Phys. Rev. B {\bf68}, 220402(R) (2003).
\bibitem[7]{Yaicle}C. Yaicle, C. Martin, Z. Jirak, F. Fauth, G. Andr\'e, E. Suard, A. Maignan, V. Hardy, R. Retoux, M. Hervieu, S. H\'ebert, B. Raveau, Ch. Simon, D. Saurel, A. Br\^ulet and Bour\'ee, Phys. Rev. B {\bf68}, 224412 (2003); C. Yaicle, F. Fauth, C. Martin, R. Retoux, Z. Jirak, M. Hervieu, B. Raveau and A. Maignan, J. Solid State Chem. {\bf178}, 1652 (2005) .  
\bibitem[8]{Ahn}K. H. Ahn, T. Lookman and A. R. Bishop, Nature {\bf428}, 401 (2004). 
\bibitem[9]{Sunil1}Sunil Nair and A. Banerjee, Phys. Rev. Lett. {\bf93}, 117204 (2004).
\bibitem[10]{Sunil2}Sunil Nair and A. Banerjee, J. Phys.: Condens. Matter {\bf 16}, 8335 (2004).
\bibitem[11]{Tokura3}T. Kimura, Y. Tomioka, R. Kumai, Y. Okimoto and Y. Tokura, Phys. Rev. Lett. {\bf83}, 3940 (1999).
\bibitem[12]{Shoji}R. Shoji, S. Mori, N. Yamamoto, A. Machida, Y. Morotomo and T. Katsufuji, J. Phys. Soc. Jpn. {\bf70}, 267 (2001).
\bibitem[13]{Mahe}R. Mahendiran, A. Maignan, S. H$\acute{e}$bert, C. Martin, M. Hervieu, B. Raveau, J. F. Mitchell and P. Schiffer, Phys. Rev. Lett. {\bf89}, 286602 (2002).
\bibitem[14]{Sharma}P.A. Sharma, S. B. Kim, T. Y. Koo, S. Guha and S-W. Cheong, Phys. Rev. B {\bf71}, 224416 (2005).
\bibitem[15]{Kumar}Kranti Kumar, A. K. Pramanik, A. Banerjee, P. Chaddah, S. B. Roy, S. Park, C. L. Zhang and S-W. Cheong, Phys. Rev. B {\bf73}, 184435 (2006); P. Chaddah, A. Banerjee and S.B. Roy, cond-mat 0601095.
\bibitem[16]{Chat}M.K. Chattopadhyay, S.B. Roy and P. Chaddah, Phys. Rev. B {\bf72}, 180401(R) (2005); P. Chaddah, Pramana-J. Phys. {\bf67}, 113 (2006).

\end{thebibliography}
\end{document}